\documentclass[sigconf,natbib=false]{acmart}

\AtBeginDocument{%
  }

\copyrightyear{2023} 
\acmYear{2023} 
\setcopyright{rightsretained} 
\acmConference[ARES 2023]{The 18th International Conference on Availability, Reliability and Security}{August 29-September 1, 2023}{Benevento, Italy}
\acmBooktitle{The 18th International Conference on Availability, Reliability and Security (ARES 2023), August 29-September 1, 2023, Benevento, Italy}\acmDOI{10.1145/3600160.3605164}
\acmISBN{979-8-4007-0772-8/23/08}

\usepackage[numbers]{natbib}
\usepackage{blindtext}
\usepackage[acronym]{glossaries}
\newacronym{tls}{TLS}{Transport Layer Security}
\newacronym{tcp}{TCP}{Transmission Control Protocol}
\newacronym{udp}{UDP}{User Datagram Protocol}
\newacronym{http}{HTTP}{Hypertext Transfer Protocol}
\newacronym{ietf}{IETF}{Internet Engineering Task Force}
\newacronym{ecn}{ECN}{Explicit Congestion Notification}
\newacronym{foss}{FOSS}{Free and open-source software}
\newacronym{dns}{DNS}{Domain Name System}
\newacronym{api}{API}{Application Programming Interface}

\usepackage{array}
\usepackage{cleveref}
\usepackage{pifont}
\usepackage{arydshln}
\newcommand{\rfcKeyWord}[1]{The RFC 2119 keyword #1 is used in the referenced section.}
\newcommand{\rfcKeyWords}[1]{The RFC 2119 keywords #1 are used in the referenced section.}

\newcommand{\checkReason}[1]{The \OK{} mark was applied if the project #1}

\usepackage{acmart-taps}

\aptLtoXcmd{
\newcommand{\Rplus}{\protect\hspace{-.1em}\protect\raisebox{.35ex}{{\textbf{+}}}}
}{
\newcommand{\Rplus}{\protect\hspace{-.1em}\protect\raisebox{.35ex}{\smaller{\smaller\textbf{+}}}}
}
\newcommand{\Cpp}{\mbox{C\Rplus\Rplus}\xspace}

\usepackage{etoolbox}
\newtoggle{submission}
\togglefalse{submission}

\newcommand*\rot{\rotatebox{77}}
\newcommand*\OK{\ding{51}}
\newcommand*\NotOK{\ding{55}}
\newcommand*\MaybeOK{[\ding{51}]}
\newcommand*\na{n/a}

\begin{document}

%%
%% The "title" command has an optional parameter,
%% allowing the author to define a "short title" to be used in page headers.
\title[A Quic(k) Security Overview]{A Quic(k) Security Overview: A Literature Research on Implemented Security Recommendations}

%%
%% The "author" command and its associated commands are used to define
%% the authors and their affiliations.
%% Of note is the shared affiliation of the first two authors, and the
%% "authornote" and "authornotemark" commands
%% used to denote shared contribution to the research.

% https://tex.stackexchange.com/a/505807
\author{Stefan Tatschner}
% \authornote{Both authors contributed equally to this research.}
\orcid{0000-0002-2288-9010}
\affiliation{%
  \institution{Fraunhofer Institute AISEC}
  \streetaddress{Lichtenbergstraße 11}
  \city{Garching bei München}
  \country{Germany}
  \postcode{85748}
}
\affiliation{%
  \institution{University of Limerick}
  \city{Limerick}
  \country{Ireland}
}
\email{stefan.tatschner@aisec.fraunhofer.de}

\author{Sebastian N. Peters}
% \authornotemark[1]
\orcid{0009-0007-6421-4023}
\affiliation{%
  \institution{Fraunhofer Institute AISEC}
  \streetaddress{Lichtenbergstraße 11}
  \city{Garching bei München}
  \country{Germany}
  \postcode{85748}
}
\email{sebastian.peters@aisec.fraunhofer.de}

\author{David Emeis}
\orcid{0009-0003-4960-944X}
\affiliation{%
  \institution{Fraunhofer Institute AISEC}
  \streetaddress{Lichtenbergstraße 11}
  \city{Garching bei München}
  \country{Germany}
  \postcode{85748}
}
\email{david.emeis@aisec.fraunhofer.de}

\author{John Morris}
\orcid{0000-0003-2811-1055}
\affiliation{%
  \institution{University of Limerick}
  \city{Limerick}
  \country{Ireland}
}
\email{john.morris@ul.ie}

% Department of Electronic and Computer Engineering, University of Limerick, Ireland; 
\author{Thomas Newe}
\orcid{0000-0002-3375-8200}
\affiliation{%
  \institution{University of Limerick}
  \city{Limerick}
  \country{Ireland}
}
\email{thomas.newe@ul.ie}

%%
%% By default, the full list of authors will be used in the page
%% headers. Often, this list is too long, and will overlap
%% other information printed in the page headers. This command allows
%% the author to define a more concise list
%% of authors' names for this purpose.
% \renewcommand{\shortauthors}{Trovato et al.}

%%
%% The abstract is a short summary of the work to be presented in the
%% article.
\begin{abstract}

Built on top of UDP, the relatively new QUIC protocol serves as the baseline for modern web protocol stacks.
Equipped with a rich feature set, the protocol is defined by a 151~pages strong IETF standard complemented by several additional documents.
Enabling fast updates and feature iteration, most QUIC implementations are implemented as user space libraries leading to a large and fragmented ecosystem.
This work addresses the research question, “if a complex standard with a large number of different implementations leads to an insecure ecosystem?”.
The relevant RFC documents were studied and “Security Consideration” items describing conceptional problems were extracted.
During the research, 13~popular production ready QUIC implementations were compared by evaluating 10~security considerations from RFC9000.
While related studies mostly focused on the functional part of QUIC, this study confirms that available QUIC implementations are not yet mature enough from a security point of view.
\end{abstract}

%%
%% The code below is generated by the tool at http://dl.acm.org/ccs.cfm.
%% Please copy and paste the code instead of the example below.
%%


%%
%% Keywords. The author(s) should pick words that accurately describe
%% the work being presented. Separate the keywords with commas.
%% A "teaser" image appears between the author and affiliation
%% information and the body of the document, and typically spans the
%% page.
% \begin{teaserfigure}
%   \includegraphics[width=\textwidth]{samples/sampleteaser}
%   \caption{Seattle Mariners at Spring Training, 2010.}
%   \Description{Enjoying the baseball game from the third-base
%   seats. Ichiro Suzuki preparing to bat.}
%   \label{fig:teaser}
% \end{teaserfigure}

%\received{30 April 2023}
%\received[revised]{30 April 2023}
%\received[accepted]{30 April 2023}

%%
%% This command processes the author and affiliation and title
%% information and builds the first part of the formatted document.
\maketitle

\section{Introduction}

QUIC is a general purpose transport layer network protocol which was designed by Jim Roskind at Google in 2012 and publicly announced as an experiment in 2013. 
While being built on top of the connectionless but efficient \gls{udp} \cite{rfc768},  it was submitted to the \gls{ietf} for standardization in 2015 and finally published as RFC9000 \cite{rfc9000}, which is supported by RFC8999 \cite{rfc8999}, RFC9001 \cite{rfc9001}, and RFC9002 \cite{rfc9002} in 2021.
As specified in RFC9001, QUIC uses the \gls{tls} \cite{rfc8446} protocol which enables authentication of peers and provides confidentiality and integrity protection for messages exchanged by endpoints.

Originally developed in the 1970s, the \gls{tcp} \cite{rfc9293} nowadays suffers from some limitations, e.g. slow connection setup times, especially when it is used with \gls{tls}.
For instance with \gls{tcp}+\gls{tls}, a client must go through both the \gls{tcp} three-way and the \gls{tls} handshake. 
QUIC was created to address some of these restrictions and in doing so, has earned it the nickname “TCP/2”.
The most common case for connection establishment requires four roundtrips for \gls{tcp}+\gls{tls} and one roundtrip for QUIC \cite{Chen2021}.

The reason for the increasing adoption of QUIC is its incorporation into the web protocol stack as
the \gls{http} is used by more than five billion people for accessing the internet \cite{statista2023}.
Facebook announced in 2020 that more than 75\% of their Internet traffic uses QUIC and HTTP/3 \cite{Chatzoglou2023}.
The current version HTTP/3 is designed to delegate features provided by the HTTP/2 framing layer to native QUIC features.
A good example are HTTP/2 \cite{rfc9113} streams which were multiplexed over a single \gls{tcp} connection in the previous \gls{http} version.
In HTTP/3, streams are provided by the QUIC protocol stack \cite{rfc9114}.

In order to allow rapid iteration of the protocol, the designers intended the protocol stack to be located in user space.
Consequently, multiple implementations for different programming languages exist today.
Being a basic building block of the current version of the most used internet protocol, different QUIC implementations are built into major browsers\footnote{\url{https://caniuse.com/http3}}, such as Chromium, Firefox, or Safari.

Due to its dissemination, QUIC carries responsibility for internet users in terms of privacy and security.
For instance, there have been multiple attempts to compromise the security of the internet.
In the past, there were reports about undersea cable tapping \cite{aldrich2020postdigital} or attempts to weaken the security parameters of \gls{tls} \cite{10.1145/3580522}.

In RFC documentation, sections called “Security Consideration” are used to discuss conceptional weaknesses in published network protocols according to the attacker model explained in RFC3552 \cite{rfc3552}.
If necessary, follow-up RFC documents are published to keep the older documents up-to-date.
Unfortunately, the current approach of the \gls{ietf} has a downside.
For instance, QUIC has a documented known vulnerability called opt-ack which is already known from \gls{tcp} and might cause a wide-spread congestion collapse \cite{10.1145/1102120.1102170}.
Countermeasures for opt-ack are well known, but for \gls{tcp}, developers decided not to implement them due to performance degradation reasons.
In the QUIC RFC, there are suggestions for a mitigation strategy but the relevant sections are tagged with the RFC2119 keywords MAY or SHOULD \cite{rfc2119} turning them into \emph{optional} countermeasures.

The research question this paper addresses is: “Does the complexity of the RFC standard on the one hand and the large number of different QUIC implementations on the other hand lead to an insecure QUIC ecosystem?”.
Initially, the conceptional weaknesses of QUIC, documented by the designers and their mitigation strategies are investigated.
Subsequently, exploratory Internet research on existing QUIC implementations is conducted.
From these results, 13 libraries were chosen for a deeper analysis.
As a next step, the selected QUIC implementations are analyzed to determine if they implemented the suggested mitigation strategies.
Finally, the results are compared and discussed.
In this study the scope was intentionally limited to the QUIC protocol, additionally incorporated protocols, such as \gls{tls}, were not considered.

The presented paper aims to give an update of the available QUIC implementations by conducting an analysis of the QUIC environment.
Further, the available analysis of vulnerabilities is complemented by comparing QUIC implementations with the security recommendations from the RFC documents.

\section{Related Work}
\label{sec:related-work}

\citeauthor{10.1145/1102120.1102170} and \citeauthor{6384958} both show in their studies from \citeyear{10.1145/1102120.1102170} and \citeyear{6384958} how misbehaving receivers could have a major impact on whole network infrastructures \cite{10.1145/1102120.1102170, 6384958}.
To overcome such emerging problems in their standards, the \gls{ietf} publishes documents updating already established RFCs.
For instance, there is RFC5961 \cite{rfc5961} specifying a modification of the \gls{tcp} inbound segment handling, reducing the chances of a successful, known attack.
On the downside, this modification is also formulated as an optional item and is not specified as a mandatory \gls{tcp} update, thereby increasing the number of possible ways of implementing the protocol.
The reason for specifying updates as optional is often for backwards compatibility.
However, there are other approaches like flag days known from the \gls{dns} which help to move the ecosystem forward\footnote{\url{https://www.isc.org/blogs/dns-flag-day-2020/}}.

\citeauthor{10.1145/3284850.3284852} created a QUIC test suite in \citeyear{10.1145/3284850.3284851} that interacts with public QUIC servers to verify their conformance with key features of the \gls{ietf} specification \cite{10.1145/3284850.3284852}.
The authors compared 15 different QUIC implementations revealing that their grade of conformance to the \gls{ietf} specification differs largely.
Additionally, this paper provides an overview of the evolution of RFC2119 keywords.
Approximately half of all tracked keywords in the QUIC specification -- at the time of this study -- were SHOULD or SHOULD NOT, respectively.
In other words, the QUIC specification contains a significant amount of optional requirements.
This work demonstrates that the complexity and optional keyword nature of the QUIC protocol leads to many implementations behaving differently for key features.

\citeauthor{10.1145/3284850.3284851} in \citeyear{10.1145/3284850.3284851} proposes a standardized logging and visualization format for debugging and analyzing QUIC \cite{10.1145/3284850.3284851}.
This proposal led to a RFC draft describing a high-level schema for a standardized logging format called qlog, which is currently at revision~5 \cite{ietf-quic-qlog-main-schema-05}.
The goal is a standardized logging format which is implemented across multiple different implementations that provides an opportunity for comparing them at runtime.

In a subsequent work, \citeauthor{10.1145/3405796.3405828} analyzed and discussed 15~different QUIC implementations in \citeyear{10.1145/3405796.3405828} \cite{10.1145/3405796.3405828}.
They used their previously presented approach based on the qlog format to debug and compare the QUIC stacks.
The authors concluded that even though these stacks all implemented exactly the same RFCs, their low-level implementation choices led to large differences in behavior between them.
These differences may be caused by the large number of optional items, as outlined in \cite{10.1145/3284850.3284852}.
It might be expected that many implementations eventually develop into a smaller set of best practices rather than the observed approaches.

In \citeyear{Chatzoglou2023}, \citeauthor{Chatzoglou2023} presented a comprehensive overview of 18 QUIC implementations and they analyzed the QUIC security through the lens of relevant literature \cite{Chatzoglou2023}.
The authors revealed several practical zero-day vulnerabilities and came to the conclusion that the available fragmented production-level implementations might not yet be mature enough.
%Unfortunately, there are some mistakes in the listing of available QUIC libraries \cite[Table 1]{Chatzoglou2023}.

\begin{table*}
\centering
    \begin{tabular}{llllll}
  \toprule
Project & $\star$ & Backed by & Language & TLS Stack & References\\
  \midrule
quic-go & 8.0k & Google & Go & modified Go stdlib & Syncthing, caddy, cloudflared, traefik \\
\hdashline
cloudflare/ \\ quiche & 7.4k & Cloudflare & Rust & Boring SSL & Android (DNS), curl, nginx \\
\hdashline
MsQuic & 3.3k & Microsoft & C & SChannel & HTTP/3 and SMB Stack on Windows\\
\hdashline
Quinn & 2.7k & Instant Domains, Inc & Rust & multiple (default: rustls) & hyper/reqwest \\
\hdashline
Neqo & 1.6k & Mozilla & C, \Cpp & NSS & Firefox \\
\hdashline
XQUIC & 1.4k & Alibaba & C & BoringSSL or BabaSSL & tengine\\
\hdashline
mvfst & 1.3k & Facebook & \Cpp & Fizz & internally at Facebook\\
\hdashline
aioquic & 1.2k & Jeremy Lainé & Python & custom & hypercorn, httpx \\
\hdashline
LSQUIC & 1.2k & LiteSpeed Technologies Inc & C & BoringSSL & internally at LiteSpeed \\
\hdashline
ngtcp2 & 0.9k & Tatsuhiro Tsujikawa & \Cpp & multiple & curl \\
\hdashline
s2n-quic & 0.9k & Amazon Web Services & Rust & s2n-tls or rustls & Amazon Web Services \\
\hdashline
quicly & 0.6k & Fastly, DeNA Co. Ltd. & C & BoringSSL & H2O webserver\\
\hdashline
google/ \\ quiche & 0.4k & Google & C, \Cpp & BoringSSL & Chromium \\
  \bottomrule
    \end{tabular}
    \caption{Overview of identified production ready QUIC implementations, sorted by Github stars. The full version of this table is available in the supplementary material.}
    \label{tab:quic-overview}
\end{table*}

\section{Methodology}
\label{sec:methodology}

This research consists of multiple, consecutive working steps.
The results were collected in separate tables including additional metadata, e.g. keywords for search or further references.
As these tables form the database for the evaluation, they will be published as additional data to this paper.
From a conceptional point of view, the methodology is structured as follows:

\begin{enumerate}
    \item \textbf{Identification of relevant RFC documents}:
        The relevant RFC documents are presented on the QUIC working group's homepage\footnote{\url{https://quicwg.org/}}.
        There are a few documents which (at the time of writing) only exist as draft documents;
        draft documents were not considered.
        During this research it turned out that only RFC9000 \cite{rfc9000} contains relevant \emph{optional} measures.
        Other documents, such as RFC9001 \cite{rfc9001}, RFC9308 \cite{rfc9308}, or RFC9312 \cite{rfc9312} contain Security Considerations but they are either not applicable to the present evaluation or they just reference RFC9000.
    \item \textbf{Analysis of “Security Considerations”}:
        The “Security Considerations” sections were analyzed in order to find concrete, documented problems with proposed solutions.
        The relevant RFC2119 keywords were extracted to outline whether the proposed solution is considered mandatory or optional.
        For the result of this work step, see Section~\ref{sec:security-considerations}.
    \item \textbf{Online search for QUIC implementations}: 
        The number of libraries for this analysis was limited by the following constraints: 
        Only implementations that consider themselves as production-ready or implementations that are already used by popular applications, such as curl\footnote{\url{https://curl.se}}.
        Only \gls{foss} projects, because an important point are publicly available discussions on development decisions.
        Support for QUIC v1. 
        The selection should cover major programming languages, including C, \Cpp, Go, Python, and Rust.
        For the result of this work step, see Table~\ref{tab:quic-overview}.
        For further reference, an extended version of this table, also including all omitted implementations, is available online\footnote{\url{https://rumpelsepp.org/projects/quic-overview}\label{foot:published}}.
    \item \textbf{Evaluation of implemented measures}:
        The identified implementations were reviewed, if the referenced ‘Security Considerations’ were implemented.
        For this purpose, a manual approach was chosen: for Github-based projects, the source code, issue tracker, and pull requests were searched for relevant keywords.
        For non Github-based projects, at least the source code was searched; additional related resources were considered if available.
        For the result of this work step, see Table~\ref{tab:quic-measures}.
        Additional information, such as the used search keywords, are published online$^{\ref{foot:published}}$.
\end{enumerate}

\section{Security Considerations}
\label{sec:security-considerations}

In their documents, the \gls{ietf} highlights possible conceptual weaknesses in standardized sections, called “Security Considerations”.
The RFC documents outlined by the QUIC working group were investigated and relevant sections were extracted.
Some text passages define optional (security related) components of the QUIC protocol that are not required by the standard to be implemented.
Indicators for optional sections are the RFC~2119 \cite{rfc2119} keywords SHOULD (NOT) and MAY (NOT);
SHOULD indicates an item is recommended, or in other words, there are valid reasons to consider a particular item.
MAY indicates that an item is truly optional.
Furthermore, there are mandatory items marked with the keyword MUST (NOT).
The following conceptional weaknesses were identified by analyzing the Security Considerations sections in the relevant RFC documents.

Table~\ref{tab:quic-measures} represents the results of this evaluation.
Each following section cites the relevant part in the RFC and gives an explanation about what the authors searched for in each implementation to add a \OK{} mark in the table.
For a few results, the authors needed to simplify the criteria, otherwise the analysis would go beyond the scope of this paper.

\subsection{Amplification}
\label{sec:amplification}

%\describedIn{Sec. 21.3}{rfc9000}.
This attack is described in \cite[Sec. 21.3]{rfc9000}.
A peer can be tricked into sending packets to a victim by abusing address validation tokens.
The proposed mitigation strategy is limiting the lifetime of the mentioned tokens.
\rfcKeyWord{SHOULD}
\checkReason{showed indications that at least basic anti-amplification mechanisms are implemented.}

\subsection{Optimistic ACK}
\label{sec:optimistic-ack}

%\describedIn{Sec. 21.4}{rfc9000}:
This attack is described in \cite[Sec. 21.4]{rfc9000}: “An endpoint that acknowledges packets it has not received might cause a congestion controller to permit sending at rates beyond what the network supports.”
The proposed strategy to detect this attack is skipping packet numbers.
\rfcKeyWord{MAY}
\checkReason{has implemented a mechanism to skip packet numbers.}

\subsection{Slowloris}
\label{sec:slowloris}

%\describedIn{Sec. 21.6}{rfc9000}:
This attack is described in \cite[Sec. 21.6]{rfc9000}:
The attacks commonly known as Slowloris  tries to keep many connections to the target endpoint open and hold them open for as long as possible.
The proposed strategy is choosing appropriate limits for a QUIC connection, for instance limiting the maximum number of clients, or a minimum transfer speed.
\rfcKeyWord{SHOULD}
\checkReason{showed indications that Slowloris is explicitly addressed or a fine-grained limiting system is available.}

\subsection{Stream Fragmentation and Reassembly}
\label{sec:stream-fragmentation}

%\describedIn{Sec. 21.7}{rfc9000}.
This attack is described in \cite[Sec. 21.7]{rfc9000}.
This attack can become arbitrary complex;
to summarize, an attacker tries to force a peer to allocate lots of memory by modifying packet numbers, for example.
The proposed mitigations could consist of avoiding over committing memory, limiting the size of tracking data structures, delaying reassembly of STREAM frames, implementing heuristics based on the age and duration of reassembly holes, or some combination of these.
\rfcKeyWord{SHOULD}
\checkReason{showed indications that one of the proposed mitigations is present or that the issue is explicitly addressed.}

\subsection{Stream Commitment}
\label{sec:stream-commitment}

%\describedIn{Sec. 21.8}{rfc9000}:
This attack is described in \cite[Sec. 21.8]{rfc9000}:
“An adversarial endpoint can open a large number of streams, exhausting state on an endpoint”.
The proposed mitigation is to set appropriate limits.
There is an own section which explains how these limits are determined \cite[Sec. 4.6]{rfc9000}.
\rfcKeyWord{MUST}
\checkReason{implemented a system to limit stream resources.}

\subsection{Peer Denial of Service}
\label{sec:peer-denial-of-service}

%\describedIn{Sec. 21.9}{rfc9000}:
This attack is described in \cite[Sec. 21.9]{rfc9000}:
“QUIC and TLS both contain frames or messages that have legitimate uses in some contexts, but these frames or messages can be abused to cause a peer to expend processing resources without having any observable impact on the state of the connection.”
The proposed technique for detecting this attack is “track cost of processing relative to progress”.
Implementations should then react with an appropriate error, for instance with terminating the connection.
\rfcKeyWords{SHOULD and MAY}
\checkReason{has implemented a system to track connection-specific statistics.}

\subsection{Explicit Congestion Notification}
\label{sec:explicit-congestion-notification}

%\describedIn{Sec. 21.10}{rfc9000}:
This attack is described in \cite[Sec. 21.10]{rfc9000}:
“An on-path attacker could manipulate the value of ECN fields in the IP header to influence the sender's rate.
RFC3168 \cite{rfc3168} discusses manipulations and their effects in more detail.”
The proposed mitigation is ignoring the \gls{ecn} field in the IP header \emph{unless} the packet passes a validation step which is described in \cite[Sec. 13.4.2]{rfc9000}.
There are no RFC2119 keywords used in this section and \gls{ecn} is an optional IP feature \cite{rfc3168}.
\checkReason{has support for the ECN field and showed indications that a validation step is implemented.}

\subsection{Stateless Reset Oracle}
\label{sec:stateless-reset-oracle}
A Stateless Reset is a response to a packet that cannot be associated with an active connection and is a possibility to stop peers continuing to send packets.
%\describedIn{Sec. 21.11}{rfc9000}:
This attack is described in \cite[Sec. 21.11]{rfc9000}:
“Stateless resets create a possible denial-of-service attack analogous to a TCP reset injection.”
This attack depends on a special condition of connection IDs and reset tokens.
Implementations must avoid this special condition.
\rfcKeyWords{MUST and MUST NOT}
\checkReason{explicitly addressed the stateless reset vulnerability.}

\subsection{Version Downgrade}
\label{sec:version-downgrade}

%\describedIn{Sec. 21.12}{rfc9000}:
This attack is described in \cite[Sec. 21.12]{rfc9000}:
“Future versions of QUIC that use Version Negotiation packets MUST define a mechanism that is robust against version downgrade attacks.“
At the time of writing, a mechanism for QUIC version negotiation exists as a \gls{ietf} draft \cite{ietf-quic-version-negotiation-14}.
A few QUIC libraries still implement draft revisions of QUIC and may be vulnerable to downgrade attacks.
\rfcKeyWord{MUST}
\checkReason{explicitly addressed the version downgrade problem - not necessarily by implementing the brand new QUIC draft already.}

% 
% https://www.rfc-editor.org/rfc/rfc9000.html#section-21.12-1
% 
% TOPIC: Future versions must implement countermeasures
% 
\subsection{Traffic Analysis}
\label{sec:traffic-analysis}

%\describedIn{Sec. 21.14}{rfc9000}:
This attack is described in \cite[Sec. 21.14]{rfc9000}:
“The length of QUIC packets can reveal information about the length of the content of those packets.”
Implementations can use PADDING frames to obfuscate the length of packets.
There are no RFC2119 keywords used in this section.
The authors are aware that hiding all information in network traffic is a complex task on its own, cf.~\cite{10.1145/2714576.2714595}.
Therefore, the \OK{} mark was applied if PADDING can be controlled via an \gls{api} from a particular application.

%%%%% ------------------------------------------------------
%%%%% EDIT tables by using this link:
%%%%%  https://fraunhofer-my.sharepoint.com/:x:/g/personal/sebastian_peters_aisec_fraunhofer_de/EVv0rEpCE75FttR2HrXhCccBW6VxGXf03viP-g0LEgeB6w?e=QOjH5O
%%%%% ------------------------------------------------------

\begin{table*}
    \centering
\begin{tabular}{lccccccccccccc}
  \toprule
Security Consideration & \rot{quic-go}   &   \rot{c/quiche}   &   \rot{MsQuic}   &   \rot{Quinn} & \rot{Neqo}  &   \rot{XQUIC*}   &   \rot{mvfst*}   &   \rot{aioquic}   &   \rot{LSQUIC}   &   \rot{ngtcp2}   &   \rot{s2n-quic}   &   \rot{quicly}   &   \rot{g/quiche*} \\
  \midrule                                                                              % quic-go   c/quiche    MsQuic      Quinn       Neqo          XQUIC       mvfst       aioquic     LSQUIC      ngtcp2      s2n-quic    quicly      g/quiche    
Amplification (cf. Sec.~\ref{sec:amplification})                                        & \OK{}     & \OK{}     & \OK{}     & \OK{}     & \OK{}       & \OK{}     & \OK{}     & \OK{}     & \OK{}     & \OK{}     & \OK{}     & \OK{}     & \OK{}     \\
\hdashline                                                                                                                             
Optimistic ACK (cf. Sec.~\ref{sec:optimistic-ack})                                      & \OK{}     & \NotOK{}  & \NotOK{}  & \NotOK{}  & \NotOK{}    & \NotOK{}  & \NotOK{}  & \NotOK{}  & \OK{}     & \NotOK{}  & \NotOK{}  & \OK{}     & \NotOK{}  \\
\hdashline                                                                                                                             
% Request Forgery (cf. Sec.~\ref{sec:request-forgery})\\                                                                               
% \hdashline                                                                                                                           
Slowloris (cf. Sec.~\ref{sec:slowloris})                                                & \OK{}     & \NotOK{}  & \NotOK{}  & \NotOK{}  & \NotOK{}    & \NotOK{}  & \NotOK{}  & \NotOK{}  & \OK{}     & \NotOK{}  & \OK{}     & \NotOK{}  & \NotOK{}  \\
\hdashline                                                                                                                             
Stream Fragmentation and Reassembly (cf. Sec.~\ref{sec:stream-fragmentation})           & \NotOK{}  & \NotOK{}  & \NotOK{}  & \NotOK{}  & \NotOK{}    & \NotOK{}  & \NotOK{}  & \NotOK{}  & \OK{}     & \NotOK{}  & \NotOK{}  & \NotOK{}  & \NotOK{}  \\
\hdashline                                                                                                                             
Stream Commitment (cf. Sec.~\ref{sec:stream-commitment})                                & \OK{}     & \OK{}     & \OK{}     & \OK{}     & \OK{}       & \OK{}     & \OK{}     & \OK{}     & \OK{}     & \OK{}     & \OK{}     & \OK{}     & \OK{}     \\
\hdashline                                                                                                                             
Peer Denial of Service (cf. Sec.~\ref{sec:peer-denial-of-service})                      & \NotOK{}  & \NotOK{}  & \NotOK{}  & \NotOK{}  & \NotOK{}    & \NotOK{}  & \NotOK{}  & \NotOK{}  & \NotOK{}  & \NotOK{}  & \MaybeOK{}& \NotOK{}  & \NotOK{}  \\
\hdashline                                                                                                                             
Explicit Congestion Notification (cf. Sec.~\ref{sec:explicit-congestion-notification})  & \na{}     & \na{}     & \OK{}     & \OK{}     & \na{}       & \na{}     & \na{}     & \na{}     & \na{}     & \OK{}     & \OK{}     & \na{}     & \OK{}     \\
\hdashline                                                                                                                             
Stateless Reset Oracle (cf. Sec.~\ref{sec:stateless-reset-oracle})                      & \NotOK{}  & \NotOK{}  & \NotOK{}  & \NotOK{}  & \NotOK{}    & \NotOK{}  & \NotOK{}  & \NotOK{}  & \NotOK{}  & \NotOK{}  & \NotOK{}  & \NotOK{}  & \NotOK{}  \\
\hdashline                                                                                                                             
Version Downgrade (cf. Sec.~\ref{sec:version-downgrade})                                & \OK{}     & \NotOK{}  & \OK{}     & \NotOK{}  & \OK{}       & \NotOK{}  & \NotOK{}  & \NotOK{}  & \OK{}     & \OK{}     & \NotOK{}  & \NotOK{}  & \OK{}     \\
\hdashline                                                                                                                             
Traffic Analysis (cf. Sec.~\ref{sec:traffic-analysis})                                  & \NotOK{}  & \NotOK{}  & \NotOK{}  & \NotOK{}  & \NotOK{}    & \NotOK{}  & \NotOK{}  & \NotOK{}  & \NotOK{}  & \NotOK{}  & \NotOK{}  & \NotOK{}  & \NotOK{}  \\
  \bottomrule
\end{tabular}
\caption{Evaluation Results. \OK{} means the mitigation is available as defined in Section~\ref{sec:security-considerations}. \MaybeOK{} means that a special case of this mitigation is implemented. \NotOK{} means the mitigation is not available. The abbreviation \na{} indicates that a required underlying feature is not available. Projects with a star * seem to be developed behind closed doors and the results need to be taken with a grain of salt. The full version of this table is available in the supplementary material.}
    \label{tab:quic-measures}
\end{table*}

\section{Evaluation}

The results of the conducted evaluation can be obtained from Table~\ref{tab:quic-measures}.
For the table, different symbols were chosen:

\begin{description}
    \item[\OK{}] The mitigation according to the defined criteria in Section~\ref{sec:security-considerations} is addressed.
    \item[\MaybeOK{}] A special case of the security consideration is implemented, therefor the defined criteria in Section~\ref{sec:security-considerations} are only partially addressed.
    \item[\NotOK{}] The mitigation according to the defined criteria in Section~\ref{sec:security-considerations} is \emph{not} addressed.
    \item[\na{}] The mitigation is out of scope and not available for the project, since a required underlying feature is not implemented.
    \item[*] The project is not developed in public but behind closed doors. Such projects have very little public information available, such as code reviews or discussions about the design decisions.
    The results of these projects need to be taken with a grain of salt.
\end{description}

Three projects, mvfst, google/quiche, and XQUIC, have public available source code but the development process itself is not visible.
The XQUIC project was very difficult to analyze with the chosen approach.
The project contains over 60k lines of C code but the code history is structured into only 78~commits.
Since there is no further documentation or metadata in the form of discussions or issues available, it was required to analyze the source code of XQUIC and rely on comments or function names.

All projects implement measures against Amplification attacks.
The security considerations Optimistic ACK (implemented by three), Slowloris (implemented by three), and Version Downgrade (implemented by six) were addressed by a few projects.
These special weaknesses were relatively easy to spot, because they were often documented in the source code or in the issue tracker.
Stream Fragmentation is addressed by LSQUIC.
The project even offers documentation about buffer handling which is claimed to be not vulnerable to stream fragmentation attacks by its authors.
Stream Commitment was addressed by all projects.
Stream Commitment was found to be similar to the Peer Denial of Service consideration that no project has addressed.
The difference between those two is that mitigating Peer Denial of Service requires a more fine grained tracking of connection resources and limits.
Quinn and quicly were the only projects that implement a fine-grained connection statistics system.
However, this statistics system is only used for meta information and not for triggering certain actions, such as terminating a connection.
There is one exception, though: s2n-quic implements a measure against reflection attacks\footnote{\url{https://github.com/aws/s2n-quic/issues/1259}}, which falls into the category of Peer Denial of Service but it is a special case.
Explicit Congestion Notification requires an underlying feature to be present.
Only four projects have implemented this feature.
If the underlying feature was implemented, the relevant security consideration was addressed as well.
Traffic Analysis is the only item, which is not addressed by any of the QUIC implementations. 

The authors of the s2n-quic chose a commendable and transparent approach.
They extracted RFC2199 keywords from all involved RFCs and created a machine readable compliance overview.
Each RFC2199 item was assigned a status (e.g. completed, missing tests, \dots) and if available a tracking issue on Github.
Unfortunately, most items relevant for this study (cf. Section~\ref{sec:security-considerations}) had status set to “unknown”.
The weblink to the compliance report is very large due to the usage of hash values in the URL.
Therefore, it is not appropriate for this paper's layout.
However, the weblink is included in the supplementary material.

During the evaluation, a reference\footnote{\url{https://github.com/ngtcp2/ngtcp2/issues/626}} to a later standard, \emph{DNS over Dedicated QUIC Connections} \cite{rfc9250} was noticed.
This standard includes a section that references RFC9000 according to traffic analysis.
This section is specified with the keyword MUST: “Implementations MUST protect against the traffic analysis attacks described in Section~7.5 by the judicious injection of padding. This could be done either by padding individual DNS messages using the EDNS(0) Padding Option (\dots) or by padding QUIC packets (\dots).” \cite[Sec. 5.4]{rfc9250}.

During the evaluation, it was further noticed that an IETF draft \emph{Compatible Version Negotiation for QUIC} \cite{ietf-quic-version-negotiation-14} is intended to be published as RFC9368 but has not yet been approved by one of its authors\footnote{\url{https://www.rfc-editor.org/auth48/rfc9368}}.
Interestingly enough, LSQUIC seems to be the first project claiming RFC9368 support.
This (potential) RFC9368 requires downgrade protection with the keyword MUST.

\section{Discussion}

According to Section~\ref{sec:methodology}, this study used a manual approach for evaluating the implemented measures.
In contrast to presented related studies in Section~\ref{sec:related-work}, there is no test suite or automated tooling available.
A manual approach is more sensitive to human error than a machine-aided approach; especially because the relevant data needs to be collected by hand from different sources, such as code comments, Github issues and documentation.
This approach does \emph{not endorse} the correctness of an implemented countermeasure, it only shows that the developers are aware of certain problems that need to be addressed.
The authors are aware that some problems, for instance Slowloris, Stream Commitment, or Peer Denial of Service, are expected to be addressed by similar looking countermeasures.
However, a manual approach enables accidental findings like RFC9250 where certain optional security considerations from RFC9000 are required with the keyword MUST.
Because the basic standard lacks keywords for Traffic Analysis, cf. Section~\ref{sec:traffic-analysis}, no project has implemented suitable \glspl{api} up to now.
Consequently, this study confirms the results presented in related work \cite{Chatzoglou2023, 10.1145/3284850.3284851, 10.1145/3284850.3284852}: The QUIC ecosystem suffers from being fragmented with many implementations that have unique differences.

It was observed that only one project implements a mitigation strategy for Stream Fragmentation (cf. Section~\ref{sec:stream-fragmentation}) and no project implements Peer Denial of Service mitigation. 
One possible explanation for this is that our search technique used is not suitable for this kind of mitigation.
Since Stream Fragmentation and Peer Denial of Service are complex topics, the text in the RFC document for the former only suggests using generic techniques, such as avoiding over-committing memory or limiting the size of data structures, rather than specifying a particular mitigation technique.
For the latter, tracking the cost for each connection is suggested and then appropriate actions should be triggered in case certain limits are reached.
Due to the generic nature of these mitigations, it is assumed that perhaps they were not recognized by the manual approach.
No project seems to address Traffic Analysis (cf. Section~\ref{sec:traffic-analysis}).
The authors have formed the impression that Traffic Analysis was not on the radar of most developers.
There are Github issues\footnote{\url{https://github.com/ngtcp2/ngtcp2/issues/626}} \footnote{\url{https://github.com/mozilla/neqo/issues/784}} asking for \glspl{api} to specify PADDING, but no one has written the required code yet.
It is expected that this will change and suitable \glspl{api} will start to appear, because the recently published RFC9250 explicitly requires the implementation of Traffic Analysis.

While related studies mostly focused on the functional part of QUIC, this study found that the hypothesis of a fragmented ecosystem is also true for the security related parts.
The RFC authors' decision to locate the QUIC protocol stack into user space, led to separate implementations from each of the big players, such as Cloudflare, Facebook, Google, and Microsoft.
For instance, Google maintains two distinct QUIC implementations: quic-go and google/quiche which is shipped with Google Chrome.

During the analysis, it was also noticed that the governance of the analyzed projects differs strongly.
There are projects like quic-go, quiche, or s2n-quic which are developed in public with a large community and a lot of searchable information on Github.
On the contrary, there are projects like mvfst, XQUIC, or google/quiche where the code is publicly available but development seems to happen behind closed doors.
This is demonstrated by the fact that there is little public information or discussion about design decisions available online. 
The mvfst and google/quiche projects use commit messages that carry extra information for internal use by Facebook or Google respectively.
In the XQUIC project, there are very few commits in the repository (currently~78), but major features with plus 12k~lines of code changes were merged without public review\footnote{\url{https://github.com/alibaba/XQUIC/pull/287}}.
Additionally, there are unanswered issues about testing\footnote{\url{https://github.com/alibaba/XQUIC/issues/265}} and concerns about easy to guess address tokens\footnote{\url{https://github.com/alibaba/XQUIC/issues/266}}.
Since the source code is used primarily for collecting information, such projects are very difficult to analyze using the chosen approach.
Nevertheless, multiple implementations help testing the robustness, standards compliance, and interoperability.

In order to improve the security, sustainability and maintainability of the QUIC ecosystem as a whole, the authors suggest creating a test suite (or extend an existing one, cf. \cite{10.1145/3284850.3284852}) which covers security consideration tests at runtime.
At best, passing these tests should become a requirement to identify as a compliant QUIC implementation.
Good examples for existing test suites in different working areas are xfstests\footnote{\url{https://git.kernel.org/pub/scm/fs/xfs/xfstests-dev.git/about/}} for filesystems and litmus\footnote{\url{http://webdav.org/neon/litmus/}} for WebDAV. 
For tracking the standards compliance, the approach of Amazon is a good example and is worth including into the proposed test suite.
Furthermore, the authors suggest proposing QUIC for inclusion in standard libraries or even operating system kernels.
Being part of a larger software project ensures long term maintenance and simplifies usage and research.
Both approaches seem to be a good and sustainable solution that is already being worked on\footnote{\url{https://github.com/golang/go/issues/44886}} \cite{10.1145/3242102.3242106}.

Despite the fact that the ecosystem is fragmented by the existence of many different but inter-operable QUIC implementations, an open question remains:
“Does the lack of implemented security considerations have an impact on the operational security of QUIC?”
A future study examining QUIC projects at runtime for actual vulnerabilities caused by the lack of particular security considerations and possible obstacles in the development might be worthwhile.

\section{Conclusion}

This research paper confirmed the results of previous studies associated with the QUIC ecosystem.
From a security point of view, the QUIC ecosystem is fragmented and no mature general purpose implementation is available.
However, there are well maintained implementations available that are used in production software, for instance quic-go is used in the Caddy web server\footnote{\url{https://caddyserver.com}}.
Big players tend to create their own implementations and ship them with their software rather than using and contributing to already available implementations. 
For inexplicable reasons these software stacks are available as Open Source software, but the development happens behind closed doors.

Answering the formulated research question: “Does the complexity of the RFC standard on the one hand and the large number of different QUIC implementations on the other hand lead to an insecure QUIC ecosystem?”, the authors would answer this question with an emphatic \emph{yes}.
The complex RFC standard with a lot of optional items and the lack of QUIC being available in major software stacks led to a large number of different implementations that addressed the same problems differently.
The study confirmed that a lot of known and documented security considerations are not addressed by most QUIC implementations.
Reducing the number of available QUIC implementations and including those into major software stacks like standard libraries or operating system kernels might help to improve the ecosystem.

\section{Data Availability}

We provide supplementary material under the CC0 (“No Rights Reserved”) license at \url{https://rumpelsepp.org/projects/quic-overview}.
Among others, the material provides more detailed versions, also including timestamps and weblinks, of Table~\ref{tab:quic-overview} and Table~\ref{tab:quic-measures}.

%%
%% The acknowledgments section is defined using the "acks" environment
%% (and NOT an unnumbered section). This ensures the proper
%% identification of the section in the article metadata, and the
%% consistent spelling of the heading.

\begin{acks}
Many thanks to Daniel and Florian from the segfault.fm podcast (\url{https://segfault.fm}) for inspiring this paper in episode 0x1a.
% \todo{is that a serious thing to write?}
%   stefan: yes :D
The authors also would like to thank Michael Heinl for providing his Overleaf account for writing this paper.
This work was partially supported by the German Federal Ministry of Education and Research~(BMBF) under Grant No. 16KIS1847 and partially by the German Federal Ministry for Economic Affairs and Climate Action~(BMWK) under Grant No. 13I40V010A.
\end{acks}

% MUST start on a new page, according to submission guidelines.
%\clearpage

%%
%% The next two lines define the bibliography style to be used, and
%% the bibliography file.
\bibliographystyle{ACM-Reference-Format}
\bibliography{bibliography}

% \printbibliography

\end{document}